\documentclass[aps,amsmath,pra,amssymb,twocolumn,showpacs]{revtex4}

\usepackage{graphicx}
\usepackage{dcolumn}
\usepackage{bm}
\usepackage{psfrag}

\begin{document}

\title{High-order harmonic generation from polyatomic molecules\\ including nuclear motion and a nuclear modes analysis}

\author{C.~B.~Madsen}
\affiliation{Lundbeck Foundation Theoretical Center for Quantum
System Research, Department of Physics and Astronomy, Aarhus University, DK-8000 Aarhus C, Denmark}

\author{M.~Abu-samha}
\affiliation{Lundbeck Foundation Theoretical Center for Quantum
System Research, Department of Physics and Astronomy, Aarhus University, DK-8000 Aarhus C, Denmark}

\author{L.~B.~Madsen}
\affiliation{Lundbeck Foundation Theoretical Center for Quantum
System Research, Department of Physics and Astronomy, Aarhus University, DK-8000 Aarhus C, Denmark}

\date{\today}
\begin{abstract}
We present a generic approach for treating the effect of nuclear motion in the high-order harmonic generation from polyatomic molecules. Our procedure relies on a separation of nuclear and electron dynamics where we account for the electronic part using the Lewenstein model and nuclear motion enters as a nuclear correlation function. We express the nuclear correlation function in terms of Franck-Condon factors which allows us to decompose nuclear motion into modes and identify the modes that are dominant in the high-order harmonic generation process. We show results for the isotopes CH$_4$ and CD$_4$ and thereby provide direct theoretical support for a recent experiment [Baker {\it et al.}, Science {\bf 312}, 424 (2006)]
that uses high-order harmonic generation to probe the ultra-fast structural nuclear rearrangement of ionized methane.
\end{abstract}


\maketitle

\section{Introduction}
In the past decade several experiments have demonstrated the potential for high-order harmonic generation (HHG) to probe molecular structure. These experiments include studies of orbital structure~\cite{itatani:nature:2004,kanai:nature:2005,itatani:PRL:2005,torres:PRL:2007}, nuclear dynamics~\cite{baker:science:2006,wagner:pnas:2006} and more recently coupled electronic and nuclear dynamics~\cite{baker:PRL:2008,li:science:2008}. The interplay between nuclear motion and HHG has been subject to several theoretical studies in the case of H$_2^+$ and H$_2$~\cite{kreibich:PRL:2001,lein:PRL:2005,chirila:PRA:2008,bandrauk:PRL:2008}  and other diatomic molecules~\cite{madsen:PRA:2006} where nuclear motion is simple in the sense that there exists only one mode of vibration (along the internuclear axis). So far, only limited theoretical work deals with the influence of nuclear motion on HHG from non-linear molecules~\cite{walters:JPB:2007,patchkovskii:PRL:2009}. 

Five years it was predicted that HHG may be used to probe the fast nuclear motion in H$_2$ by comparing the HHG spectrum to that of the isotope D$_2$~\cite{lein:PRL:2005}, and this was soon after confirmed in a pioneering experiment~\cite{baker:science:2006}. The D$_2$/H$_2$ ratio of the harmonic spectra is an increasing function of the harmonic order which can be understood from the basic idea that the harmonic order is associated with the time the electron spends in the continuum from initial ionization to recombination, $\tau$. The larger $\tau$, the higher the harmonic order. At 800 nm and intensities around $10^{14}$ W/cm$^2$ typical $\tau$'s are in the 1-2 fs regime and the reason for the observed increase is the faster nuclear motion in the lighter isotope H$_2$, leading to a smaller overlap of the nuclear wave packets in the recombination step. In the experiment~\cite{baker:science:2006} similar results were reported for HHG from methane isotopes (CD$_4$/CH$_4$).
 
The current work addresses the effect of nuclear motion on HHG for arbitrary, linear or non-linear,  polyatomic molecules. As we will see, the effect is generally incorporated through a nuclear correlation function describing the nuclear wave packet dynamics in the molecular cation from initial ionization to recombination. This function is highly demanding to determine. To solve the problem, we first express the nuclear correlation function in terms of Franck-Condon (FC) factors (defined as the square of the overlap integral between the vibrational wavefunctions of the neutral molecule and the molecular cation, in their respective electronic states) and the accompanying time-dependent phases caused by the vibrational excitations. Second, we consider these in the harmonic approximation, where they may be calculated using standard approximations and technology from quantum chemistry.
The theory is exemplified by calculations on CH$_4$ and CD$_4$. Methane has nine modes for vibrational relaxation, but the model used here allows us to identify the two most important vibrational modes which turn out to promote a T$_d$ $\leftrightarrow$ C$_{2v}$ nuclear geometry reconfiguration [see, e.g., Ref.~\cite{atkins:book:2005} for a discussion of point group symmetry]. We thereby prove a conjecture put forward in~\cite{baker:science:2006}.

This paper is organized as follows. In Sec.~\ref{sec:theory}, we present the theory. The electronic part is well-known from other works, so we focus on the analysis of the vibrational motion. In Sec.~\ref{sec:results},  we present results on CH$_4$ and CD$_4$. Section \ref{sec:conclusion} concludes. 
In Appendix~\ref{app:FCfactors} we discuss the calculations of the FC factors.
Atomic units [$\hbar = e = m_e= a_0 = 1$] are used throughout this paper.

\section{Theoretical model and computational details}
\label{sec:theory}

We wish to provide a simple model that in a general and transparent way isolates the role of nuclear motion in the typical experiment on HHG from molecules in the gas phase. We treat electronic dynamics in the Lewenstein model~\cite{lewenstein:PRA:1994} following the implementation of Ref.~\cite{torres:PRL:2007}, but adapt a few improvements to the latter as detailed below.

The observable quantity is the HHG spectrum, which we calculate from the Fourier transform of the dipole velocity $\langle\bm{v}(t)\rangle$~\cite{diestler:PRA:2009}, .i.e., 
\begin{equation}\label{eq:spectrum}
  S(\omega)=\left(\frac{1}{\omega T}\left\vert\int_0^T dt\, e^{-i\omega t}\langle \bm{v}(t)\rangle\right\vert\right)^2,
\end{equation}
where the laser field driving the process is non-zero only in the time interval $[0,T]$. Since the early 1990's it has been an ongoing discussion whether the HHG spectrum should be calculated from the (pulse limited) Fourier transform of the dipole moment, the dipole velocity or the dipole acceleration. For relatively long and weak laser pulses the final result is independent of the choice: Up to a well-known frequency dependent factor one can interchange the dipole velocity with the dipole moment or the dipole acceleration, since the appropriate boundary terms vanish almost exactly. For short and intense laser pulses this is no longer true, and the use of a wrong form leads to an unphysical background in the HHG spectrum~\cite{burnett:PRA:1992,bandrauk:PRA:2009}. We have tested our calculations using the different forms, and we find good agreement between the velocity and the acceleration form, whereas the length form differs considerably.
These findings are consistent with previous studies~\cite{chirila:JModOpt:2007}, and so we restrict our analysis in this paper to the velocity form results.

To evaluate the dipole velocity $\langle\bm{v}(t)\rangle=\langle\Psi(t)\vert\bm{v}\vert\Psi(t)\rangle$ in Eq.~\eqref{eq:spectrum} we apply the Born-Oppenheimer approximation, the strong-field approximation and single-active-electron model. We further freeze all orbitals except the highest occupied molecular orbital (HOMO) and consider the orientation of the nuclei as fixed during the short high-harmonic generating femtosecond pulse $\bm{F}(t)=F(t)\bm{e}$ of linear polarization $\bm{e}$. 
Then introducing the molecular orbital of the active electron evaluated at the nuclear equilibrium configuration, $\Psi_0(t)=\psi_0(\bm{r})\exp(iI_pt)$  ($I_p$ is the adiabatic ionization potential of the molecule), the ground vibrational state of the neutral molecule, $\chi_{i,0}$, and using a product of a Volkov wave, $\psi^V_{\bm{k}}$, and the $\nu$th vibrational state of the molecular ion, $\chi_{f,\nu}$ for the propagator we have
\begin{align}
\vert\Psi(t)\rangle&=\vert\Psi_0\chi_{i,0}(t)\rangle\nonumber
\\&-i\int_0^t dt'\int d^3k\sum_\nu\vert\psi^V_{\bm{k}}(t)\chi_{f,\nu}(t)\rangle\nonumber
\\&\times\langle\psi^V_{\bm{k}}(t')\chi_{f,\nu}(t')\vert F(t')\bm{e}\cdot\bm{r}\vert\psi_0(t')\chi_{i,0}(t')\rangle
\end{align}
and thus
\begin{align}\label{eq:dipolev}
\langle\bm{v}(t)\rangle&=i\int_0^tdt'\,C(t-t')F(t')\nonumber\\
&\times\int d^3k\bm{v}_{\text{rec}}^*(\bm{k}+\bm{A}(t))d_{\text{ion}}(\bm{k}+\bm{A}(t'))e^{-iS(\bm{k},t,t')}\nonumber\\
&+c.c.,
\end{align}
where 
\begin{align}
\label{C1}
C(t-t')= \sum_{\nu} \exp\left[-i\epsilon_{\nu}(t-t')\right]\left\vert\langle\chi_{f,\nu}\vert\chi_{i,0}\rangle\right\vert^2
\end{align}
 is the vibrational autocorrelation function (Fig.~\ref{fig:vibPrinciple}) with $\vert\langle\chi_{f,\nu}\vert\chi_{i,0}\rangle\vert^2$ FC factors (see Appendix~\ref{app:FCfactors}), $\epsilon_\nu$ the vibrational energy, 
\begin{align}
\bm{v}_{\text{rec}}(\bm{k})&=\bm{k}(2\pi)^{-3/2}\int d^3r\exp[-i\bm{k}\cdot\bm{r}]\psi_0(\bm{r})\nonumber\\&\equiv\bm{k}\phi_0(\bm{k}),\\ 
d_{\text{ion}}(\bm{k})&=\bm{e}\cdot(2\pi)^{-3/2}\int d^3r\exp[-i\bm{k}\cdot\bm{r}]\bm{r}\psi_0(\bm{r})\nonumber\\&=i\bm{e}\cdot\bm{\nabla}_{\bm{k}}\phi_0(\bm{k}),\\
S(\bm{k},t,t')&=\int_t'^tdt''[(\bm{k}+\bm{A}(t''))^2/2+I_p],
\end{align}
and $\bm{A}(t)$ the vector potential of the laser field.

The electronic part of Eq.~\eqref{eq:dipolev} has appeared many times since the seminal paper~\cite{lewenstein:PRA:1994} and clearly points out the three essential steps of HHG process: The electron
ionizes to the continuum at time $t'$ with probability amplitude $F(t')d_{\text{ion}}(\bm{k}+\bm{A}(t'))$. It then
propagates in the field until time $t$ acquiring a phase factor $S(\bm{k},t,t')$ and recombines with a probability amplitude $\bm{v}_{\text{rec}}^*(\bm{k}+\bm{A}(t))$. Due to vibration this product of amplitudes is weighted by a nuclear factor $C(t-t')$~\cite{lein:PRL:2005}. 

To evaluate the electronic part of Eq.~\eqref{eq:dipolev} we write the molecular orbital of the active electron in the molecular fixed (MF) frame as a linear combination of Gaussian orbitals %
that we find using the GAMESS quantum chemistry code~\cite{GAMESS}. Our basis choice is TZV with polarization functions. The expansion of the molecular orbitals in terms centered on the atoms allows us to evaluate the $k$-integrals of Eq.~\eqref{eq:dipolev} within the improved stationary phase method~\cite{chirila:PRA:2006,faria:PRA:2007,adam:PRA:2010}, where we include electron trajectories leading from one atomic center to another. Dipole velocities for different orientations of the molecule can be obtained by applying the Euler rotation operator~\cite{zare:book:1988,torres:PRL:2007} to the molecular fixed wave function.

\begin{figure}
\centering
\includegraphics[width=1.0\columnwidth]{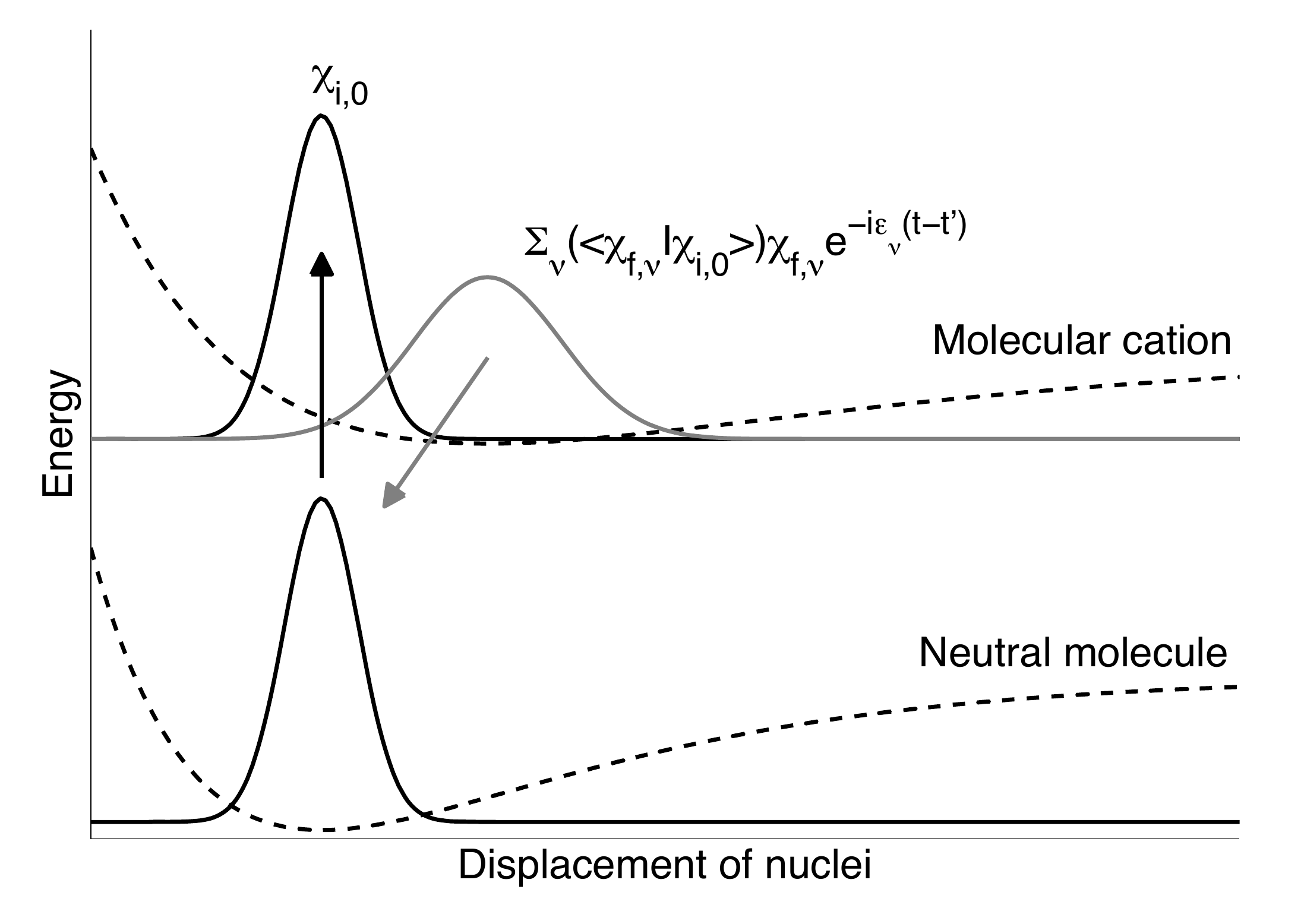}
\caption{(Color online) Simple sketch of the effect of vibration on HHG. When the molecule ionizes the laser launches a vibrational wave packet that evolves on the ionic Born-Oppenheimer surface. The overlap of this wave packet and the initial vibrational state [see Eq.~\eqref{C1}] weights the dipole velocity [see Eq.~\eqref{eq:dipolev}] and hence the HHG signal.}\label{fig:vibPrinciple}
\end{figure}

\section{Results and Discussion}
\label{sec:results}

\begin{figure}
 \centering               
 \includegraphics[width=1.0\columnwidth]{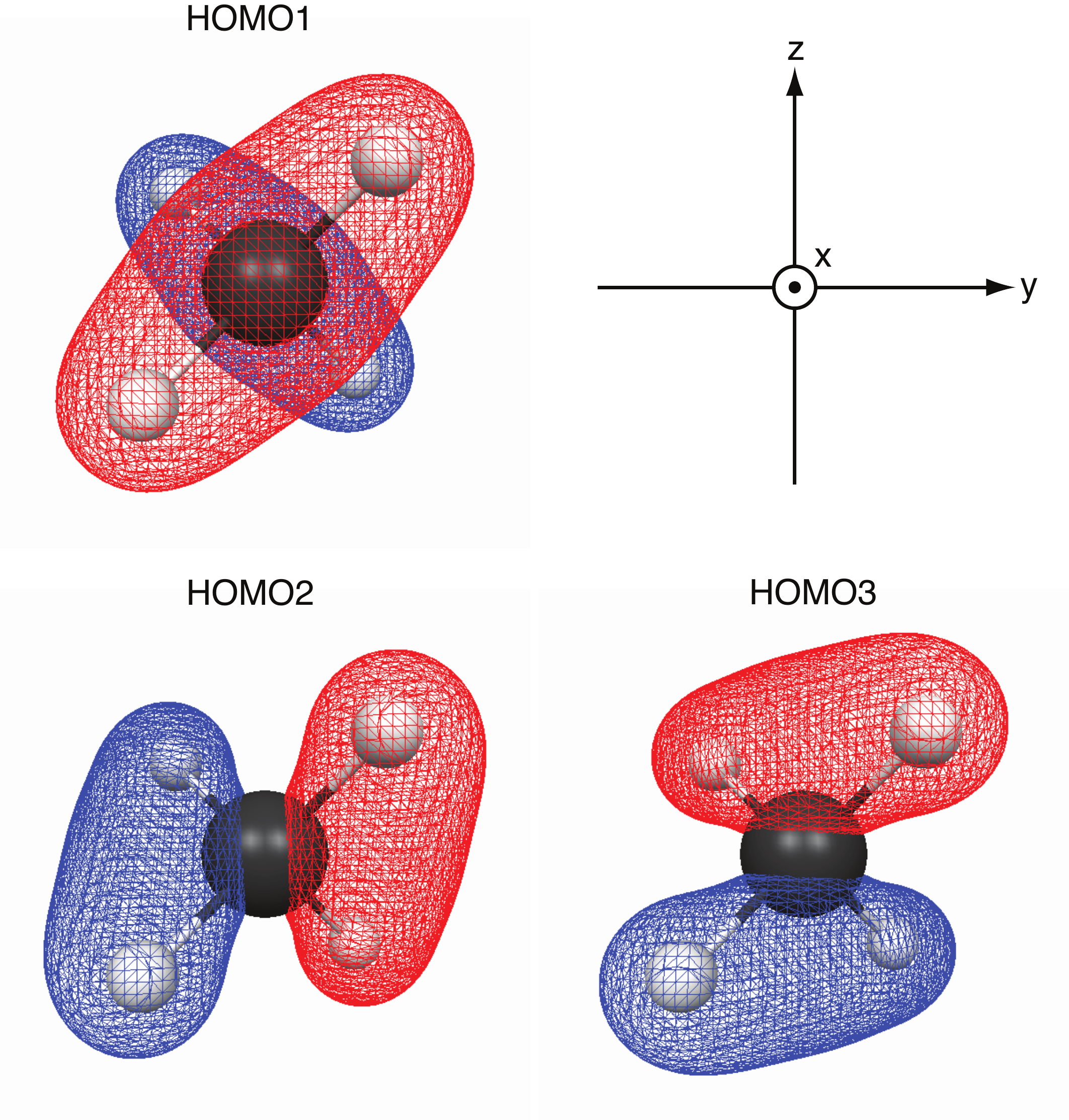}
 \caption{(Color online) Isocontour plots of the degenerate HOMOs of methane. We calculate HHG yields with the linear laser field polarization along the $x$ axis and include only the contribution from HOMO1.}\label{fig:orbitals}
 \end{figure}

\begin{figure}
\centering
\includegraphics[width=1.0\columnwidth]{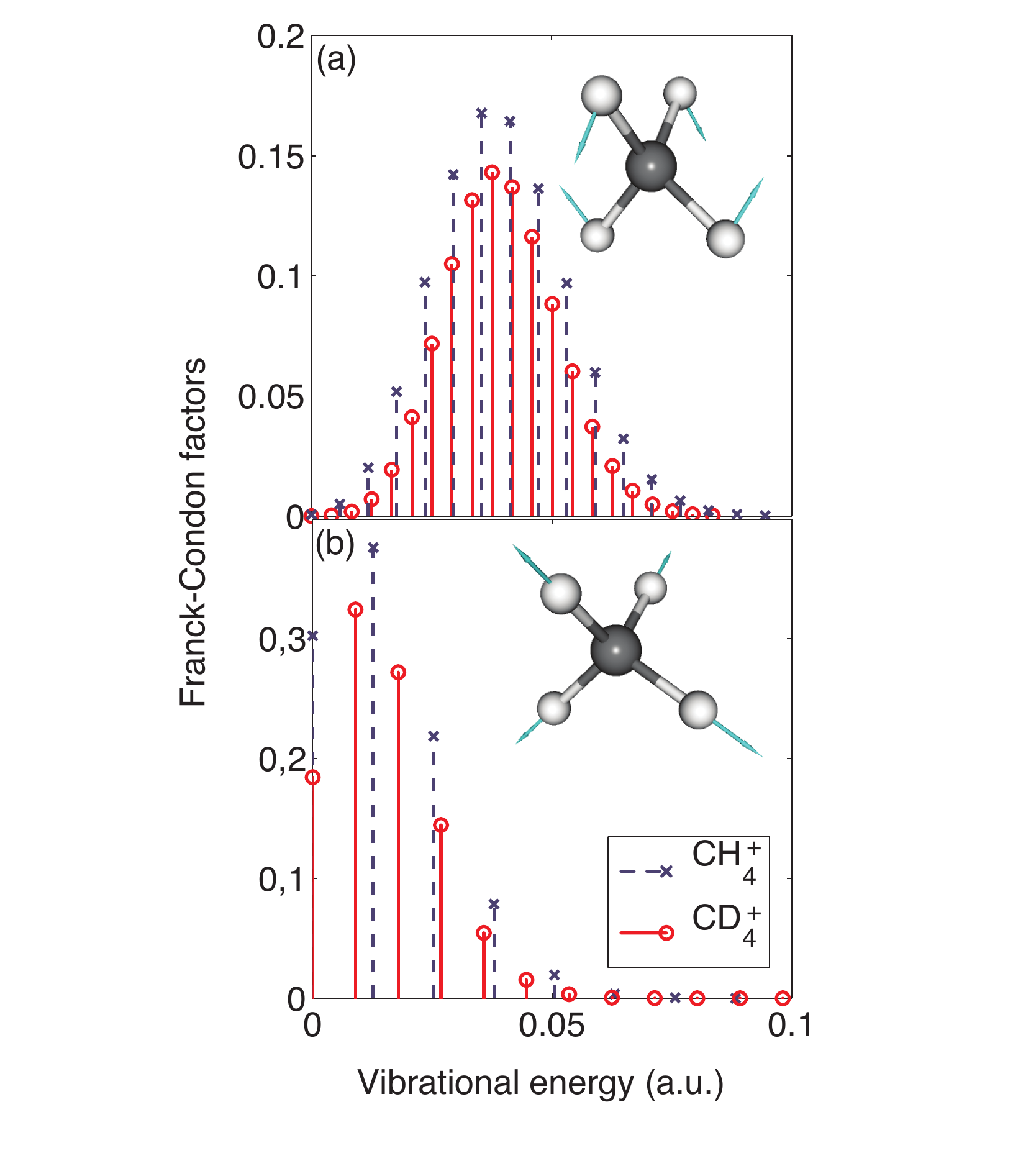}
\caption{(Color online) Franck-Condon factors for the E mode (panel (a)) and A$_1$ mode (panel (b)) of CH$_4^+$ (dashed) and CD$_4^+$ (full). These two dominating modes will drive the molecule from the T$_d$ symmetry into the relaxed C$_{2v}$ symmetry upon ionization as conjectured in~\cite{baker:science:2006}.}
\label{fig:FCFs_2}
\end{figure}

We now turn to an application of the theory described above. We consider HHG from CH$_4$ and CD$_4$. To calculate the harmonic yields, we first determine the electronic structure and the FC factors. Using GAMESS we find the following configuration of the molecular orbitals $(1a_1)^2(2a_1)^2(3t_2)^6$ and thus we have six electrons distributed among three degenerate HOMOs (see Fig.~\ref{fig:orbitals}) that could all contribute appreciably to the harmonic yield. 
The adiabatic ionization potential ($I_p=12.92$ eV) is estimated by comparing the total energy of the methane to that of the relaxed methane ion. The effective ionization potential is then obtained by adding the additional shifts ($\epsilon_\nu$) due to vibrational excitation.
Our FC analysis shows that only two normal modes are excited when CH$_4$ (CD$_4$) ionizes. These are an E symmetry mode ($\omega\approx$1295~cm$^{-1}$ for CH$_4^+$; $\omega\approx$920~cm$^{-1}$ for CD$_4^+$) that brings the molecule towards a plane and  an A$_1$ symmetry mode ($\omega\approx$2766~cm$^{-1}$ for CH$_4^+$; $\omega\approx$1960~cm$^{-1}$ for CD$_4^+$) that correspond to changes of the C$-$H (C$-$D) bond lengths. In Fig.~\ref{fig:FCFs_2} we show the one-dimensional FC factors for both modes (calculated from integrals over a single mode as shown in Appendix~\ref{app:FCfactors}) along with insets indicating the nuclear rearrangements related to each normal mode. These modes drive the molecular ion into the relaxed C$_{2v}$ symmetry.

With the molecular structure at hand, i.e., molecular orbitals and FC factors, we should, in principle, calculate the molecular dipole velocity as given in Eq.~\eqref{eq:dipolev} for different orientations of the molecule and for each of the degenerate orbitals and average the resulting dipole velocities [see Ref.~\cite{madsen1:PRA:2007}]. However, due to the high symmetry of the methane molecule there is only a small angular dependence of the harmonic yield and further the yield is dominated by the contribution from a single orbital that couples strongly to the linearly polarized laser field. We have verified this by calculations for various orientations not shown here. Consequently, the averaging is redundant and the results presented in this paper are carried out for one fixed orientation of the molecule and includes only the yield of the dominant orbital (cf. Fig.~\ref{fig:orbitals}). 
Another technical detail is that we limit the $t'$ integral in Eq.~\eqref{eq:dipolev} to times when $\tau=t-t'$ is smaller than 0.65 times an optical cycle. We do this to have a one-to-one correspondence between the time the electron spends in the continuum (from the instant ionization, $t'$, to the moment of recombination, $t$) and the harmonic order, i.e., the energy released when the electron recombines~\cite{lein:PRL:2005}.

\begin{figure}
   \centering
   \includegraphics[width=1.0\columnwidth]{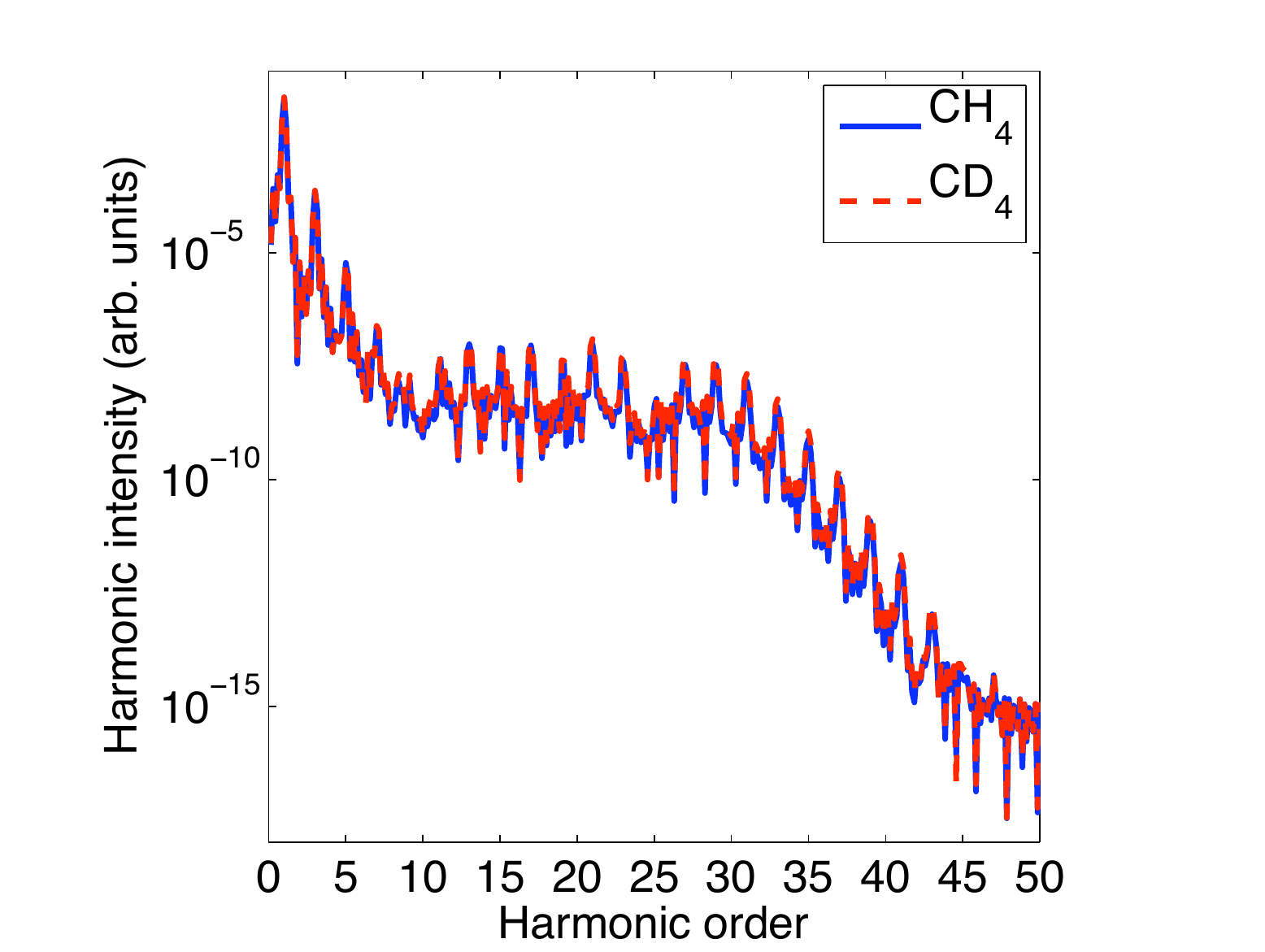}
   \caption{(Color online) Harmonic spectra for CH$_4$ and CD$_4$ using a 775 nm linearly polarized laser of intensity $2\times10^{14}$ W/cm$^2$ and the trapezoidal envelope detailed in the main text.
}\label{fig:spectrumAndWavelet}
 \end{figure}

Figure~\ref{fig:spectrumAndWavelet} shows the harmonic spectra of CH$_4$ and CD$_4$. We use a trapezoidal shape for the vector potential $\bm{A}(t)=A(t)\hat{\bm{e}}_x$, with two optical cycles linear turn-on and turn-off and three cycles of constant amplitude corresponding to peak intensity of $2\times10^{14}$ W/cm$^2$. The carrier wavelength is 775 nm. The curves shown on the figure exhibit the typical characteristics of harmonic spectra, viz., an exponential drop-off at low harmonic orders followed by a plateau with a cutoff around the harmonic order 33 in agreement with the cutoff formula~\cite{lewenstein:PRA:1994}. 

\begin{figure}
   \centering
   \includegraphics[width=1.0\columnwidth]{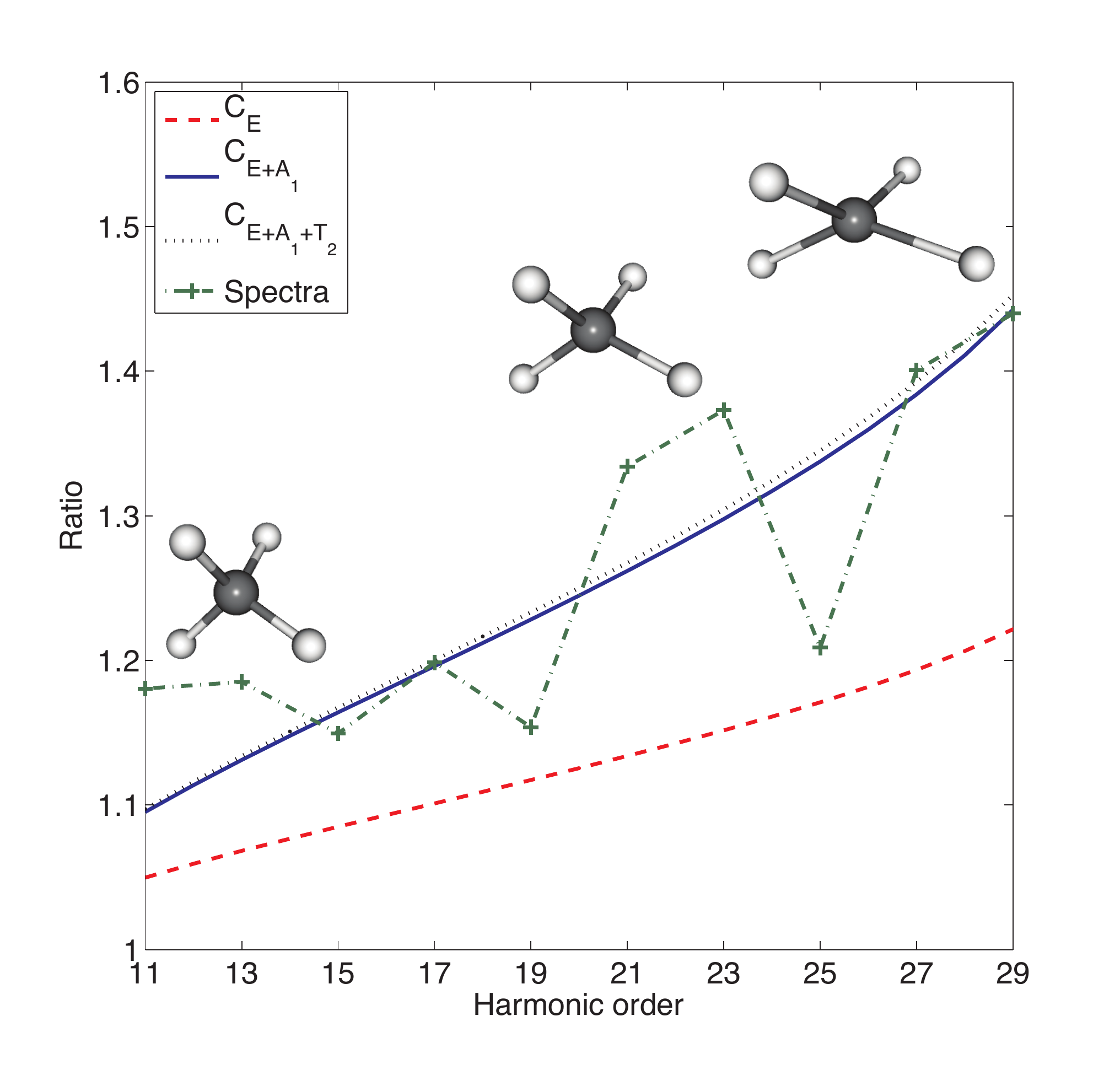}
   \caption{(Color online) Curves displaying the ratio of the harmonic spectra from Fig.~\ref{fig:spectrumAndWavelet} (dash-dotted) and of the CD$_4$/CH$_4$ nuclear correlation functions (Eq.~\eqref{C1}) including normal modes of symmetry E (dashed), E and A$_1$ (full) and E and A$_1$ and T$_2$ (dotted). 
}\label{fig:CD4vsCH4}
 \end{figure}

It is hard to see the difference in the spectra for the two isotopes, but if we integrate the spectra in an interval around each (odd) harmonic order and plot the ratio of these numbers for the isotopes, we get the result shown in Fig.~\ref{fig:CD4vsCH4} by the dash-dotted curve. For comparison, we also plot the CD$_4$/CH$_4$ ratio of the nuclear correlation functions (see Eq.~\eqref{C1}) if we include only the dominant E symmetry mode ($C_{\text{E}}$), both the E and the A$_1$ symmetry mode ($C_{\text{E+A$_1$}}$) and if we further include two slightly excited $T_2$ symmetry modes ($C_{\text{E+A$_1$+T$_2$}}$). 
We see that the E+A$_1$ motion accounts for the major effect of nuclear motion. Further, the nuclear correlation functions are increasing monotonically and hence in our model the oscillatory structure of the HHG ratio arises from the combined electron-nuclear dynamics.


\begin{figure}
 \centering               
 \includegraphics[width=1.0\columnwidth]{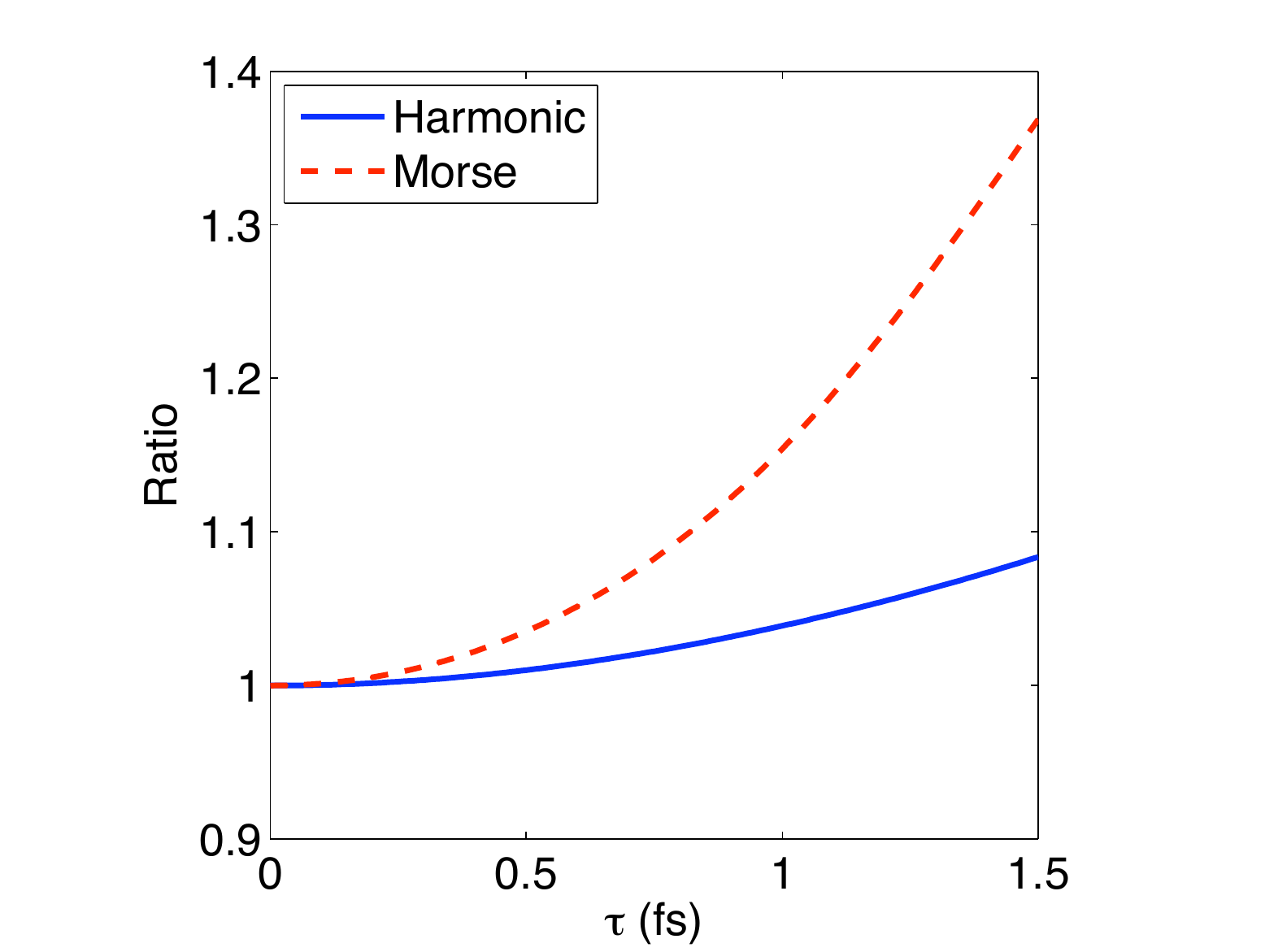}
 \caption{(Color online) Ratio of the D$_2$/H$_2$ nuclear correlation function (Eq.~\eqref{C1}) for the harmonic approximation (solid) and the Morse potential (dashed).}\label{fig:harmVsMorse}
 \end{figure}

The ratio predicted by the current theory underestimates the slope of the ratio as compared to measurements~\cite{baker:science:2006}. To understand this deviation we refer to more detailed calculations carried out on the simpler systems H$_2$ and D$_2$. For these systems we may clearly understand the consequences of the approximations made here, and as such attribute the disagreement to two factors. First, we expect that the harmonic approximation, used here to retrieve the FC factors, yields nuclear factors, $C(t-t')$, with a too low ratio between the isotopes since stretching of the molecule is underestimated when the asymmetry of the potential is not taken into account. To substantiate this conjecture we have checked the case of H$_2$  and D$_2$, where we can compare the nuclear correlation functions resulting from FC factors based on the harmonic approximation and a more accurate Morse potential~\cite{dunn:JCP:1966}, respectively and our reasoning is validated (see Fig.~\ref{fig:harmVsMorse}).
Second, we have not included any coupling of nuclear and electron dynamics. In HHG from the isotopes H$_2$ and D$_2$ such coupling is known to result in a dynamic two-center interference effect that leads to a higher ratio in the D$_2$/H$_2$ spectrum~\cite{baker:PRL:2008}.

The nuclear correlation function is expressed in terms of the FC factors (Eq.~\eqref{C1}). To this end we note that for systems where the weight on the different FC factors are effectively experimentally adjustable by laser pulse preparation one may control the nuclear correlation function. This is for instance the case in H$_2$~\cite{calvert:JPB:2008} and we have checked that by populating only FC factors $(0,\nu=0)$ and $(0,\nu=18)$ the nuclear correlation function oscillates with a period of $1.56$ fs.  Consequently, the nuclear correlation function can be modulated within less than half an optical period of the typical HHG driving field (1.33 fs for an $800$ nm field) and thereby drive enhancement and suppression of certain electron trajectories and associated harmonic orders. Control of the relative strength of the harmonics is for instance useful for attosecond pulse generation~\cite{farkas:pla:1992} and the control scheme briefly discussed here is based on the intrinsic structure of the molecule. Finally, since the nuclear correlation function is expressed in terms of one-dimensional FC factors, we can identify the important part of the nuclear dynamics by studying the changes of the nuclear correlation function as we gradually add the different normal modes and by including the minimal amount of modes in the full HHG calculation we save CPU time.

\section{Conclusion and outlook}
\label{sec:conclusion}
In conclusion, we have followed~\cite{lein:PRL:2005} and applied the Lewenstein model~\cite{lewenstein:PRA:1994} to molecules with moving nuclei. We assume that the electronic and nuclear part separate. The electronic part is then treated conventionally. For the nuclei, however, 
 we relate the vibrational autocorrelation function to FC factors [see Eq.~\eqref{C1}] and associated dynamical phase factors. For some polyatomic molecules the FC factors and energies are available in the literature, and the vibrational part of the theory can be determined directly.

In the cases where FC factors and vibrational energies are not available, we discuss how to perform a normal mode analysis and calculate the intensity of the FC transitions in the harmonic approximation. The model covers any polyatomic molecular system where the electron dynamics is reasonably described within the single-active-electron picture.
The theoretical and computational models involved are fairly standard within strong-field physics and computational chemistry. 
Finally, with the present theory, we identify the vibrational modes involved in the ultrafast rearrangement of the nuclei in the CH$_4$/CD$_4$ system, and obtain qualitatively agreement with the measurements~\cite{baker:science:2006}.

\section{Acknowledgements}
This work is supported by the Danish Research Agency (Grant no. 2117-05-0081). C.B.M. acknowledges the hospitality of the Physics Department at Kansas State University, USA, where part of this work was performed.  

\appendix

\section{Computation of the nuclear factor of Eq.~\eqref{C1}}
\label{app:FCfactors}

We are interested in evaluating Franck-Condon integrals, i.e., $\langle \chi_{f,\nu}\vert \chi_{i,0}\rangle$,
between a vibrationally-cold initial state ($\chi_{i,0}$) and a vibrationally-excited final state ($\chi_{f,\nu}$)
where the subscript $\nu$ denotes the excitation level.

In the normal-coordinate representation within the harmonic approximation, the vibrational wavefunctions describing
the initial state (i.e., the neutral molecule) and the final states (i.e., the ion), respectively, are expressed
as
\begin{equation}
\chi_{i,0}({\bf Q^\prime}) = \left(\det {\bm \Gamma^\prime}/\pi^{N_m}\right)^{\frac{1}{4}}  
\exp{\left(-\frac{1}{2}{\bf Q^\prime}^\dagger {\bf \Gamma^\prime} {\bm Q^\prime} \right)},
\end{equation}
and
\begin{align}
\chi_{f,{\bm n}}({\bf Q}) &= \left(\det {\bm \Gamma}/\pi^{N_m}\right)^{\frac{1}{4}}
\exp{\left(-\frac{1}{2}{\bf Q}^\dagger {\bf \Gamma} {\bm Q} \right)} 
\nonumber\\&\times\prod_j^{N_m} \left(2^{n_j} {n_j}!\right)^{-1/2}H_{n_j}({\bm \Gamma}^{1/2}{\bm Q}),
\end{align}
where ${\bm Q}=(Q_1, Q_2, \cdots, Q_{N_m})$ is a normal coordinate vector, ${\bm \Gamma}$ is a diagonal
matrix with elements $\Gamma_{j,j}=\hbar/\omega_j$ where $\omega_j$ is the vibrational
frequency of mode $j$, $N_m$ is the number of vibrational
degrees of freedom, the index ${\bm n}=(n_1,n_2,\cdots,n_{N_m})$ is a vector of vibrational excitations ($\bm{n}$ is related to $\nu$ through the function $A$ that orders the excited modes according to ascending energy, i.e., $\nu(\bm{n})=A(\bm{n})$) and $H_{n_j}$ is the $n_j$th Hermite polynomial.
For a vibrationally-cold state,  ${\bm n}=(0_1,0_2,\cdots,0_{N_m})\equiv0$.
The normal coordinates of the initial state (${\bm Q^\prime}$) and the final state (${\bm Q}$) are related by
a simple transformation~\cite{Toniolo2001}, i.e.,
\begin{equation}
\label{AppEq3}
{\bf Q^\prime} = {\bf J}{\bf Q} + {\bf \Delta},
\end{equation}
where ${\bf J}$ is the so-called Duschinsky matrix and the vector ${\bf \Delta}$ expresses the geometry change in the final state.
The ${\bf J}$ matrix reflects the mapping of the normal coordinates of the initial-state onto those of the final-state.

The multi-dimensional Franck-Condon integral, within the harmonic approximation, reads
\begin{align}
\label{AppFCint}
\langle \chi_{f,{\bm n}}\mid \chi_{i,0}\rangle&=N \int dQ_1\ldots dQ_{N_m} H_{n_1}(\Gamma_1Q_1)\ldots \nonumber\\&\times H_{n_{N_m}}(\Gamma_{N_m}Q_{N_m}) \nonumber\\ 
&\times \exp{ \left[ -\frac{1}{2}\Gamma_1Q_1^2   - \cdots - \frac{1}{2}\Gamma_{N_m}Q_{N_m}^2 \right]}\nonumber\\&\times  
\exp{ \left[ -\frac{1}{2}\Gamma_1^\prime {Q_1^\prime}^2  
- \cdots - \frac{1}{2}\Gamma_{N_m}^\prime {Q^\prime_{N_m}}^2 \right]}.
\end{align}
where the normalization factor is given by
\begin{equation}
N =\prod_j^{N_m} \left( \frac{\Gamma_j^{1/2}{{\Gamma_j^{\prime}}^{1/2}}}{\pi2^{n_j}{n_j}!} \right)^{1/2}.
\end{equation}

When evaluating the integrals, a considerable simplification is introduced by assuming that the off-diagonal elements in
${\bf J}$ (see Eq.~\eqref{AppEq3}) are very small, i.e.,
\begin{align}
Q_j^\prime &= J_{j,1} Q_1 + J_{j,2} Q_2 + \cdots + J_{j,N_m}Q_{N_m} + \Delta_j \nonumber\\&\approx J_{j,1} Q_1 + \Delta_j.
\end{align}

Accordingly, the multi-dimensional Franck-Condon integral reduces to a product of one-dimensional integrals,
i.e,
\begin{align}
\langle \chi_{f,{\bm n}}\mid \chi_{i,0}\rangle &=  \prod_{j}^{N_m} \int dQ_j H_{n_j}(\Gamma_jQ_j)\nonumber\\& \times \exp\left[ -\frac{1}{2}\Gamma_jQ_j^2  -\frac{1}{2}\Gamma_j^{\prime} (Q_j-\Delta_j)^2  \right].
\end{align}

The Ansbacher recurrence relations are used to obtain one-dimensional FC integrals~\cite{Ansbacher}.
Computing FC integrals involves calculation of the equilibrium geometries and normal modes for the neutral and the ionized molecules. These were obtained from calculations 
using the hybrid density functional B3LYP level of theory in conjunction with the triple-$\zeta$ valence basis set as implemented in Gaussian~\cite{g03}.


\end{document}